# TESTS OBSERVATIONNELS CLEFS DE LA COSMOLOGIE MODERNE

**Monique Signore**
Observatoire de Paris, LERMA
61, avenue de l'Observatoire, 75014 Paris, France
Monique.Signore@obspm.fr
&
**Denis Puy**
Institut de Physique Théorique
Université de Zürich , 8057 Zürich, Suisse
puy@physik.unizh.ch

L'expansion de l'Univers a été une des rares théories astrophysiques qui a précédé son observation. Lorsque Alexandre Friedmann et l'Abbé Lemaitre modifièrent indépendamment l'un de l'autre les équations d'Einstein (décrivant la structure de l'Univers) et tirèrent l'existence d'une expansion universelle, aucune observation ne suggérait cet élément essentiel à la compréhension de l'évolution de l'Univers. Ce n'est que, 20 années plus tard, qu'Edwin Hubble mis en évidence la récession des galaxies, prouvant ainsi que l'Univers n'est pas statique.

Si l'Univers est en expansion, la première réflexion fondamentale est qu'il a été beaucoup plus petit et dense par le passé. Partant de cette idée simple, on peut imaginer que l'Univers *naquit* à partir d'une singularité extremement dense et chaude. C'est ainsi que Fred Hoyle -partisan d'un Univers statique- railla cette idée dans une émission scientifique de la BBC en introduisant le vocable désormais célèbre: le *Big Bang*.

Depuis lors, de nombreux faits observationnels directs et indirects permirent d'affiner cette approche et confirmèrent cet incroyable résultat: l'Univers se dilue !

Jusqu'à il y a encore une dizaine d'années, le modèle classique de l'expansion constante de l'Univers était pour la grande majorité des astrophysiciens une théorie très satisfaisante.

Néanmoins un énorme problème se posa aux théoriciens à propos de la formation des grandes structures telles que amas de galaxies et galaxies. En effet dans un Univers se dilatant lentement depuis son origine, des régions spatiales éloignées sont *causalement* déconnectées: aucun signal ou information, se déplacant obligatoirement à une vitesse inférieure à la vitesse de la lumière n'a pu se propager de l'une à l'autre. En fait un rapide calcul montre que les hétérogénéités de matière (prises comme germes des futures structures) sont non-causales entre elles. Ce point, nommé le *problème de l'horizon*, était particulièrement insatisfaisant pour une théorie complète de l'Univers: un contact causal était nécessaire.

Alan Guth imagina un audacieux mécanisme: l'inflation. Il introduisit un champ scalaire: l'inflaton, dont les particules associées sont dotées d'une masse (c'était possible ici, étant données les gigantesques énergies de l'Univers primordial). Ainsi, si on considère une région de l'espace primordial suffisamment grande, l'évolution temporelle du facteur d'échelle de cette région est alors gouvernée par le contenu énergétique de l'inflaton (cinétique et potentiel). Très rapidement (au premier instant de l'Univers), l'énergie de l'inflaton se réduit à sa seule énergie potentielle. L'évolution de la région sera alors régie par cette énergie potentielle qui change soudain spectaculairement: son facteur d'échelle croit exponentiellement. Il en résulte que toutes les distances , qui croissent comme ce facteur d'échelle, deviennent plus grandes que le *rayon* de l'Univers à cette époque. La géométrie de l'espace s'aplatit alors, les régions deviennent *causales*. A cause de cette nouvelle expansion, l'inflaton va peu à peu connaitre une fin brutale:
il se désintègre en photons (on parle de réchauffement dans ce cas précis) puis l'évolution rejoint celle du Big Bang chaud. De récentes observations, comme nous allons le voir, viennent, non seulement, de confirmer les prédictions de cette théorie, mais aussi de suggérer que l'Univers est plus *léger* que prévu et que son *expansion est en train d'accélérer.*

**1- Les mesures du fond de rayonnement cosmologique et la géométrie de l'Univers**

A l'époque primordiale - i.e. post-inflatoire- la matière et le rayonnement étaient intrinséquement liés par les processus de diffusion, et plus précisément par le processus de diffusion Thomson des photons sur les électrons et les charges libres.

La dilution de la matière va engendrer, comme nous l'avons vu, toute une série de transitions de phase. Par exemple une très importante transition pour la cosmologie est la transition quark-hadrons ou le confinement des quarks en hadrons. A cette transition, les hadrons vont peu à peu interagir en vue de former les premiers noyaux de l'Univers (essentiellement d'hydrogène, de deuterium, d'helium, de lithium et quelques traces de bore et de beryllium). Ces noyaux vont alors se combiner avec les électrons ambiants afin de produire les atomes correspondants. Cette décroissance électronique aura des conséquences fatales sur le couplage thermique entre la matière et le rayonnement. Peu à peu matière et rayonnement se découpleront, *libérant* alors les photons et produiront un rayonnement baignant l'Univers dans son ensemble.

Cette perte d'opacité du rayonnement fut définitivement effective lorsque l'Univers eut un *age* d'environ 300 000 ans (pour mémoire, son *age* actuel est approximativement de 15 milliards d'années). Ces photons fossiles (ou rayonnement de fond cosmologique), observés maintenant aux fréquences micro-onde et radio, ont une énergie moyenne qui correspond à celle d'un spectre d'un corps noir à T=2.728 Kelvins [1]. De façon encore plus remarquable, si l'on *balaie* le ciel, on peut constater de petites variations de température -au niveau de 1 sur 100 000- (voir figure 1).

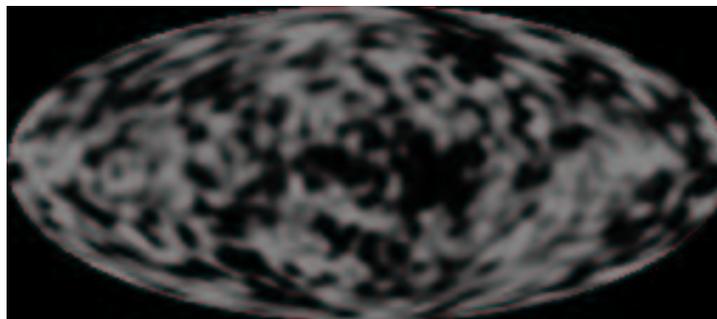

Figure 1: Carte des mesures du rayonnement de fond cosmologique obtenue par le satellite COBE en 1992 [1]-

Ces fluctuations nous fournissent un instantané des hétérogénéités de rayonnement qui existaient à l'époque du découplage entre la matière et le rayonnement, et par conséquent des indications sur les conditions variables de la distribution de matière à cette époque.
Le modèle inflatoire se révèle alors extrémement précieux, car il offre une prédiction détaillée et très spécifique de ces *taches chaudes* et *froides* sur le ciel. Il prédit combien il doit y avoir de taches de telle ou telle taille angulaire sur le ciel, et de combien ces taches sont plus chaudes ou plus froides que la température moyenne.

Ces endroits chauds (ou froids) de la carte sont dus aux photons qui ont émergé des régions les plus (ou moins) denses à l'époque du découplage thermique entre la matière et le rayonnement. La taille angulaire de ces taches se révèle etre alors la prédiction la plus importante du modèle inflatoire. En fait la taille caractéristique de ces régions peut se calculer à partir des équations d'équilibre gravitation-pression à l'époque du découplage.
La relation entre la taille physique et la taille angulaire apparente observée sur le ciel dépend, de facon cruciale, de la géométrie de l'Univers. Si la partie spatiale de la métrique est euclidienne (on parle alors d'Univers plat) les taches sous-tendent un angle d'environ 1° sur le ciel -une courbure spatiale négative rendrait la taille apparente sur le ciel plus petite, à l'opposé d'une courbure spatiale positive qui la donnerait plus grande.

Il y a un peu plus d'un an, les expériences BOOMERANG et MAXIMA (ballons mesurant le fond de rayonnement) donnèrent une confirmation éclatante des prédictions du modèle inflatoire. Ces deux ballons obtinrent le résultat fantastique suivant: les taches les plus chaudes et les plus froides sont exactement de la taille angulaire correspondant à une géométrie spatiale euclidienne, géométrie justement prédite par l'inflation ! (voir figure 2).

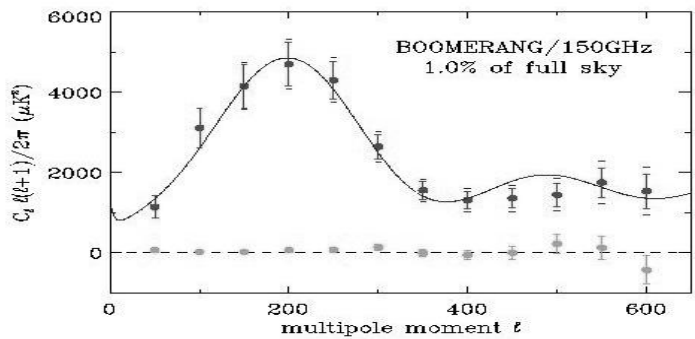

Figure 2: Fluctuations de la température du rayonnement de fond cosmologique en fonction de la taille angulaire obervée par BOOMERANG [2]. *La position du premier pic du spectre dépend de la géométrie de l'Univers. Ces observations montrent clairement que le premier pic apparait pour une taille angulaire de 1° ( i.e. l=200), position prédite par le modèle inflatoire, celle d'un Univers plat. Dans le cas d'une Univers ouvert (courbure spatiale négative) le pic serait sur la gauche du pic observé. Dans un Univers fermé (courbure spatiale positive) il serait sur la droite.*

## 2-Sur les mesures de la masse de l'Univers et la matière noire

Dans le cadre du modèle classique du Big Bang et de l'inflation, en mesurant le taux d'expansion, on peut prédire la densité de matière à partir de la Relativité Générale comme étant approximativement de $10^{-29}$ grammes par centimètre cube. Cette valeur est connue comme étant la densité critique -i.e. celle qui correspond à une géométrie plate de l'Univers- $\rho=3H/8\pi G$ ou $H$ est la constante de Hubble (liée au taux d'expansion) et G est la constante de gravitation, Notons ici qu'en général, on introduit la lettre grecque $\Omega$ pour désigner le rapport de la densité moyenne à la densité critique de l'Univers. Jusqu'à ces dernières années, avec la confirmation observationnelle de ses prédictions, on pensait que l'inflation avait maitrisé la géométrie (plate), déterminé la densité ($\Omega=1$) et décidé, en quelque sorte, du destin de l'Univers: à jamais une expansion à un taux toujours en décélération.

Pourtant, il faut noter que, bien avant cette confirmation observationnelle de l'inflation, en faisant l'inventaire de toutes les formes de la matière dans l'Univers, et en calculant ainsi sa densité totale on avait trouvé qu'elle était bien inférieure à la densité critique: $\Omega_m<<1$. En effet, depuis des décennies, on sait que la somme de toute la matière ordinaire ou baryonique (i.e. neutrons et protons) contribue seulement très faiblement à la valeur critique correspondant à un Univers plat. En d'autres termes, la matière serait essentiellement d'un type différent de la matière baryonique. En fait, depuis les années 1930, de nombreuses mesures ont indiqué qu'il doit exister une matière invisible -ou matière noire- dans l'Univers pour expliquer, en particulier, les courbes de rotation des galaxies dans les amas de galaxies.

Cette *matière noire* pourrait etre constituée d'une composante de *matière exotique* i.e. de nouvelles particules élémentaires telles qu'axions, neutralinos, higgsinos ... -et d'une plus faible composante de *matière ordinaire*: astéroides, naines brunes, trous noirs, nuages moléculaires froids ect... Pendant longtemps on a pensé que la densité totale de la matière (*baryonique* et *noire*) pourrait atteindre la densité critique : $\Omega_m=\Omega_{bar}+\Omega_{noire}=1$.

Mais, ces dernières années, avec les résultats de diverses mesures par des méthodes indépendantes -en particulier celles de l'évolution des amas de galaxies- il apparait dans la communauté des astrophysiciens un consensus selon lequel la densité totale de la matière (matière baryonique et matière noire) serait inférieure à la moitié de la densité critique (i.e. $\Omega_m<1/2$).

## 3- Les mesures de l'expansion et l'énergie noire.

Mais alors, comment est-il possible que la densité de matière soit inférieure à la moitié de la densité critique ($\Omega_m<1/2$) et que l'Univers soit plat ($\Omega=1$) ?

La relativité Générale serait-elle fausse ? Bien sur que non ! Selon la théorie d'Einstein, considérons l'Univers présent, composé de matière ordinaire et de matière noire, d'une courbure et d'énergie du vide (i.e. la fameuse constante cosmologique $\Lambda$ introduite par Einstein lui-meme). Les contributions relatives de la densité de matière ($\rho_m$), de la densité du vide ($\rho_\Lambda$) et de la courbure sont données respectivement par $\Omega_m=8\pi G\rho_m/3H$ ; $\Omega_\Lambda=8\pi G\rho_\Lambda/3H$ et $\Omega_k$. L'équation de Friedmann est alors donnée par : $\Omega_m+ \Omega_\Lambda + \Omega_k=1$.

Les mesures du fond de rayonnement nous apprennent que l'Univers est plat ($\Omega_k=0$), les mesures de la masse de la matière noire nous suggèrent que $\Omega_m<0.5$ et l'équation de Friedmann nous conduit à $\Omega_m+ \Omega_\Lambda =1$.

Notons que si l'introduction de $\Lambda$ fut jugée obsolète -après la découverte de l'expansion par Hubble- le concept lui-meme l'est beaucoup moins maintenant: $\Omega_\Lambda$ peut etre regardée comme une autre forme d'énergie: *l'énergie du vide* ou plus généralement *l'énergie noire*.

D'ailleurs, dans la perspective moderne de la physique des particules, le problème n'est pas l'existence de $\Lambda$, mais pourquoi $\Lambda$ n'est-il pas beaucoup plus grand ?
Dans la perspective de la cosmologie le problème est que si $\Lambda$ existe, $\Lambda$ introduit une force répulsive, une sorte d'anti-gravité qui pourrait accélérer l'expansion.
C'est justement là, le très beau résultat de 1998 de deux groupes - le *Supernova Cosmology Project* et le *High z Supernova Search*- qui, à partir de mesures précises et indépendantes de l'expansion cosmique ont montré que l'expansion de l'Univers est en train de s'accélérer.
Leur résultat repose sur la découverte de l'existence d'une relation entre le maximum de luminosité d'une supernova de type Ia et le taux avec lequel sa brillance décroît. Cette relation permit ainsi d'adopter ces objets comme des chandelles cosmiques. Rappelons que les supernovae sont des explosions stellaires; les supernovae de type Ia se distinguent des autre types par l'absence d'hydrogène et la présence de raies caractéristiques dans leur spectre émis à leur maximum de luminosité; dans la suite, on ne considère que les supernovae de type Ia qu'on désigne sous le terme générique de supernovae.
Ayant étudié la luminosité d'une supernova en mesurant sa *courbe de lumière* (i.e comment la brillance observée varie avec le temps), on détermine la distance physique à la supernova en utilisant le flux de photons observés. Alors en comparant la distance avec le redshift (qui nous dit de combien la supernova s'éloigne de nous) on peut reconstruire l'histoire de l'expansion. C'est exactement comme les bornes kilométriques que l'on voit défiler depuis un train en mouvement: le taux avec lequel les bornes arrivent vers nous et s'éloignent de nous peuvent nous donner le taux avec lequel se meut le train.
Les résultats des mesures des deux équipes sont similaires : les supernovae observées les plus lointaines sont environ 20% moins lumineuses ou encore 10% plus lointaines qu'elles ne le seraient dans un Univers en expansion constante (voir figure 3).
Depuis 1998, les astrophysiciens recherchent les erreurs possibles qui pourraient confondre les résultats: présence de poussière sur les lignes de visée, variations des supernovae elles-memes. Mais jusque là, la recherche d'artefacts s'est révélée négative et on peut conclure que notre Univers semble etre dans une phase accélérée de son expansion [4].

**4- Conclusions**

Il semble que l'observation des supernovae lointaines soit une des dernières observations-clefs de la cosmologie moderne. En combinant les mesures effectuées sur le fond cosmologique, sur les amas et sur les supernovae, on peut réunir les différentes contraintes correspondantes sur $\Omega_m$ et $\Omega_\Lambda$ -voir figure 4.
Entre 1998 et 2000 de nombreuses méthodes indépendantes conduisent à une concordance remarquable des nombres clefs qui décrivent notre Univers. Il semble que l'Univers est plat. Mais son contenu est fait d'un mélange d'ingrédients étranges. Les atomes ordinaires (baryons) des étoiles, des nébuleuses, du gaz intergalactique diffus fournissent à peu près 5% de l'énergie de la matière, cette dernière fournissant environ 25% du contenu énergétique total de l'Univers, et l'énergie noire le reste, soit 75%. L'expansion s'accélère car l'énergie noire (avec une pression négative) est le constituant dominant. De tous les atomes de l'Univers, la moitié est dans les galaxies et le reste est diffus à travers l'espace intergalactique. Les étoiles et les gaz dans les galaxies font moins que 2% du budget total masse/énergie de l'Univers.

Pour expliquer l'accélération de l'Univers, les cosmologistes modernes ont ainsi introduit *l'énergie noire* ou la *quintessence*. Ce nom a des précédents historiques:

-En philosophie, la quintessence se réfère au cinquième élément -après l'air, la terre, le feu et l'eau-proposé par les Grecs anciens pour décrire une substance sublime et parfaite.

-En littérature, Rabelais introduisit, dans Gargantua, la quintessence comme la Reine d'une science spéculative.

Désormais il est nécessaire de donner un sens au mot *quintessence* en Cosmologie. Comme une nouvelle cosmologie observationnelle émerge, on doit répondre à plusieurs défis:
- Quelle est la physique sous-jacente à l'inflation ?
- Quelles sont les particules élémentaires qui constituent l'essentiel de la matière noire ?
- Quelle est la nature de l'énergie noire ou quintessence ?

De nombreux programmes sont prévus -et certains meme sont déjà en place- tels que MAP [5], Planck Surveyor [6], SNAP [3]. Les progrès qu'on doit observer dans la prochaine décennie nous montreront si la cosmologie entre alors effectivement dans son age d'or.

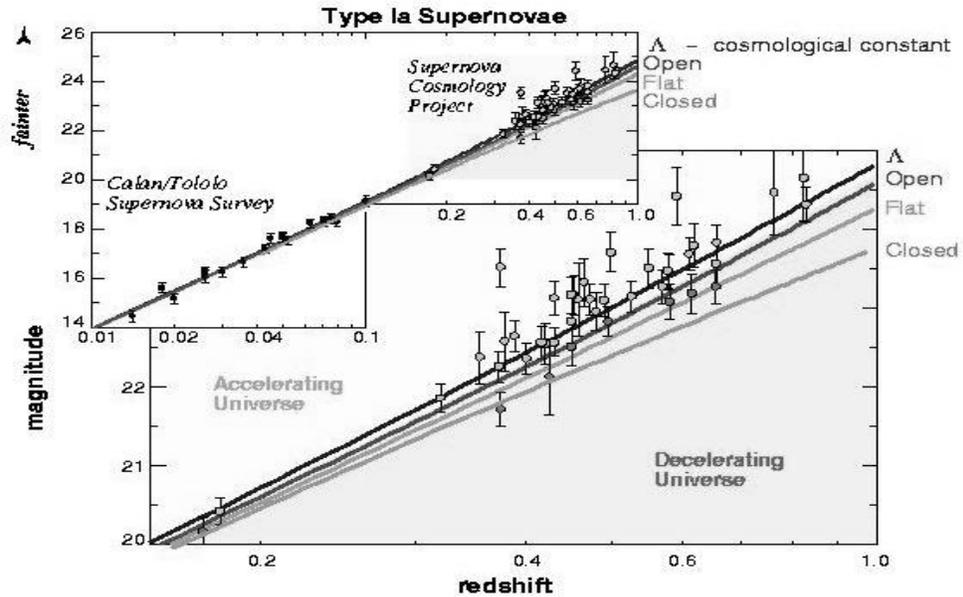

**Figure 3: Supernovae comme chandelles cosmiques (d'après Perlmutter [3]).**
La brillance observée, en unités logarithmiques de magnitude, des supernovae en fonction de leur redshift (les plus brillantes sont en bas, les moins brillantes sont en haut). Les supernovae proches ont été observées par le Telescope de Calan-Tololo au Chili. Les plus lointaines ont été observées par l'équipe du *Supernova Cosmology Project*. On peut comparer ces observations aux prédictions de plusieurs modèles théoriques. L'attraction gravitationnelle exercée dans les modèles *open* ($\Omega_m=0.28$, $\Omega_\Lambda=0$) et *flat* ($\Omega_m=1$, $\Omega_\Lambda=0$) décélèrent le taux d'expansion et rend les supernovae lointaines (i.e. à grand redshifts) apparemment plus brillantes. En revanche, dans le cas du modèle $\Lambda$ ($\Omega_m=0.28$, $\Omega_\Lambda=0.72$) l'existence d'une constante cosmologique $\Lambda$ accélère le taux d'expansion, les supernovae sont apparamment moins brillantes.

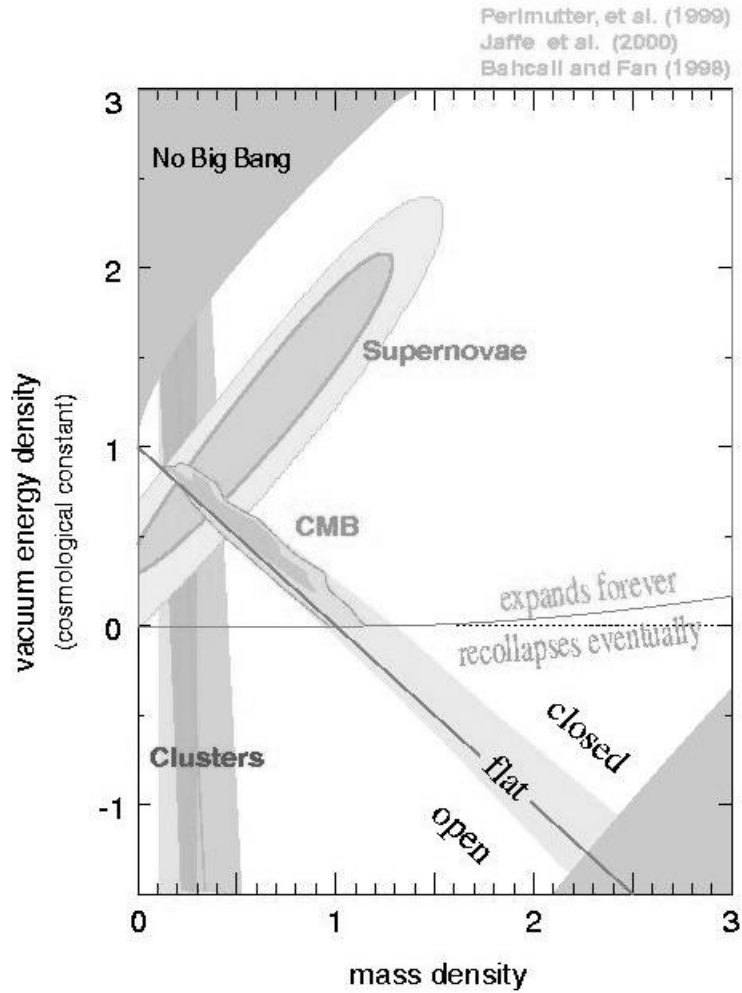

**Figure 4: Les contraintes resultant des trois measures (Fond cosmologique, Amas et Supernovae) dans le plan ($\Omega_m$, $\Omega_\Lambda$)**

Les contraintes des amas de galaxies (clusters) indiquent un Univers à faible densité; celle des supernovae indiquent un Univers en accélération alors que celles du fond cosmologique (CMB) indiquent un Univers *plat* ($\Omega_m + \Omega_\Lambda = 1$). L'intersection des trois régions suggérent que l'Univers est plat avec $\Omega_m = 1/3$ et $\Omega_\Lambda = 2/3$. Le modèle d'Univers doit impliquer l'existence d'une constante cosmologique non nulle, c'est à dire l'existence d'une *énergie noire*, i.e. la *quintessence* -d'après Perlmutter [3]. Sur ce schéma, on distingue aussi 3 régions: région médiane: Big Bang et expansion; région du haut à gauche: pas de Big Bang et expansion; région du bas à droite: Big Bang et éventuellement effondrement (Big Crunch).